\documentclass[11pt,a4paper]{iopart}
\usepackage{cite}
\usepackage{graphicx} 
\usepackage{xcolor}

\usepackage[utf8]{inputenc}

\usepackage{iopams}

\expandafter\let\csname equation*\endcsname\relax

\expandafter\let\csname endequation*\endcsname\relax 

\usepackage{amsmath}
\usepackage{amssymb}
\usepackage{hyperref}

\makeatletter
\def\@mkboth#1#2{}
\newlength\appendixwidth
\newcommand{\patchl@section}{%
  \settowidth{\appendixwidth}{\textbf{Appendix }}%
  \addtolength{\appendixwidth}{1.5em}%
  \patchcmd{\l@section}{1.5em}{\appendixwidth}{}{\ddt}%
}

\makeatother

\begin{document}

\title{Dynamically emergent correlations in bosons via quantum resetting}

\author{Manas Kulkarni}
\address{International Centre for Theoretical Sciences, Tata Institute of Fundamental Research,
Bangalore 560089, India\\
Email: manas.kulkarni@icts.res.in }

\author{Satya N. Majumdar}
\address{LPTMS, CNRS, Universit\'e Paris-Sud, Universit\'e Paris-Saclay, 91405 Orsay, France\\
Email: satya.majumdar@universite-paris-saclay.fr}

\author{Sanjib Sabhapandit~\footnote{Author to whom any correspondence should be addressed.}}
\address{Raman Research Institute, Bangalore 560080, India\\
Email: sanjib@rri.res.in}

\date{\today}

\begin{abstract}
 We study the nonequilibrium stationary state (NESS) induced by quantum resetting of a system of $N$ noninteracting bosons in a harmonic trap. Our protocol consists of preparing initially the system in the ground state of a harmonic oscillator centered at $+a$, followed by a rapid quench where the center is shifted to $-a$ and the system is allowed to evolve unitarily up to a random Poissonian time $\tau$ distributed via $r\, e^{-r \tau}$. Then the trap center is reset to $+a$ again and the system is assumed to cool instantaneously to the initial ground state.  The system is again allowed to evolve unitarily in the trap centered at $-a$ up to a random time, and the procedure is repeated. Under repeated resetting, the system reaches a NESS where the positions of bosons get {\em strongly correlated} due to simultaneous resetting induced by the trap. We fully characterize the steady state by analytically computing several physical observables
such as the average density, extreme value statistics, order and gap statistics, and also the distribution of the number of particles in a region $[-L,L]$, known
as the full counting statistics (FCS). In particular, we show that in the large $N$ limit, the scaling function describing the FCS exhibits a
striking feature: it is supported over a nontrivial finite interval, and moreover is discontinuous at an interior point of the support. Our results are supported by numerical simulations.
This is a rare example of a strongly correlated quantum many-body NESS where various observables can be exactly computed.
\end{abstract}
\newpage
\noindent\rule{\hsize}{2pt}
\tableofcontents
\noindent\rule{\hsize}{2pt}

\maketitle

\section{Introduction}

Quantum gases have been a subject of intense theoretical and experimental studies over several decades. In particular, a gas of interacting bosons, realized in cold atom systems, has been extensively studied \cite{DGPS99,BBDBG08,LLFS12}. Despite a plethora of progress, exact analytically tractable models for correlated bosonic gases are hardly available. For bosonic systems, two commonly studied models are the Lieb–Liniger model~\cite{LL63,LL63b,S2004} and the Gross-Pitaevski equation~\cite{PS08,PS03,ESY07}. Lieb–Liniger model describes a Bose gas with Dirac delta interaction in one spatial dimension. This quantum model is integrable, in the sense that its solution can be written in the form of a Bethe ansatz. Although several aspects of the  Lieb–Liniger model have been extensively studied~\cite{CC07, CC07_PRL, PC14, NP16}, especially given its integrability property, explicit solutions of various observables, such as the extreme value and the order statistics, the full counting statistics (FCS) have remained elusive. Furthermore, the inevitable presence of external confining traps (mostly harmonic) in experimental setups breaks the integrability,  making calculations even more challenging. Another well-studied model that often describes the collective behaviour of a weakly interacting Bose gas is the Gross-Pitaevski equation. This is a continuum one-dimensional model that is integrable.  Despite its integrability, it is a nonlinear partial differential equation, and therefore, exact solutions for observables remain challenging. Once again, in the presence of a confining trap, this integrable structure is broken 
making computations even more evasive~\cite{DGPS99}.  Thus, there is a growing need to engineer and study correlated Bose gas which is also analytically tractable, in addition to being experimental feasible. It is hence natural to explore experimentally feasible correlated particle systems for which several observables can be computed analytically. 

Recently, a new type of correlated gases has been found in classical systems where the correlations between particles are not inbuilt but are rather induced by the dynamics itself, e.g., by simultaneous resetting of independent Brownian particles~\cite{BLMS_23}. Despite the presence of strong correlations in the stationary state, these models are solvable because of a special structure of the joint probability density function (JPDF) in the stationary state, namely, the conditionally independent and identically distributed (CIID) structure~\cite{BLMS_23}
\begin{equation}
  P_\mathrm{st}(x_1,x_2,\dotsc,x_N) = \int_{-\infty}^{\infty} du\,  h(u) \prod_{j=1}^N p(x_j|u). 
  \label{eq:jointPDF}
\end{equation}
The JPDF $ P_\mathrm{st}(x_1,x_2,\dotsc,x_N)$ in \eqref{eq:jointPDF} is clearly not factorizable, thus rendering the gas correlated. However, inside the integral, for a fixed value of the parameter $u$, there is a factorizable structure. One can interpret this as follows. For a fixed $u$, the $\{x_i\}$ variables are independent, each drawn from a probability density function (PDF) $p(x|u)$ parametrized by $u$. But the parameter $u$ itself is a random variable distributed via the PDF $h(u)$. Once integrated over $u$, the JPDF loses the factorizabality. 
The advantage of the JPDF having a CIID structure is as follows. 
For an ideal gas without interaction (but with a certain parameter $u$ fixed), several physical observables can be easily computed analytically and one then needs to average over the parameter $u$ drawn from its distribution $h(u)$ as in \eref{eq:jointPDF}. In a series of recent works, this CIID structure \eref{eq:jointPDF} in the stationary state has been found in several classical models in one dimension~\cite{BLMS_23,BLMS24,BMMS24,SM24}.

Stochastic resetting ~\cite{EM11, EM12,ESG20, GJ22,KR24} has emerged as a major area of research in statistical physics in recent times. While most of these studies focused on resetting in classical systems, there have been relatively fewer studies on quantum resetting~\cite{MSM18,RTLG18,  PCML21,DDG21,PCL22,MCPL22,TDFS22,DCD23,JV23,YB23,KM23,KM23a,YWB24}. In this paper, we study a quantum gas of noninteracting bosons in a trap subjected to 
stochastic resetting of the many-body quantum state to the initial state. We find that this quantum system reaches a nonequilibrium stationary state (NESS) that also has
the above CIID structure \eref{eq:jointPDF}, thus allowing exact calculation of several observables. This thus presents a much-needed new solvable strongly correlated quantum gas. 

\begin{figure}
    \centering
    \includegraphics[width=.8\linewidth]{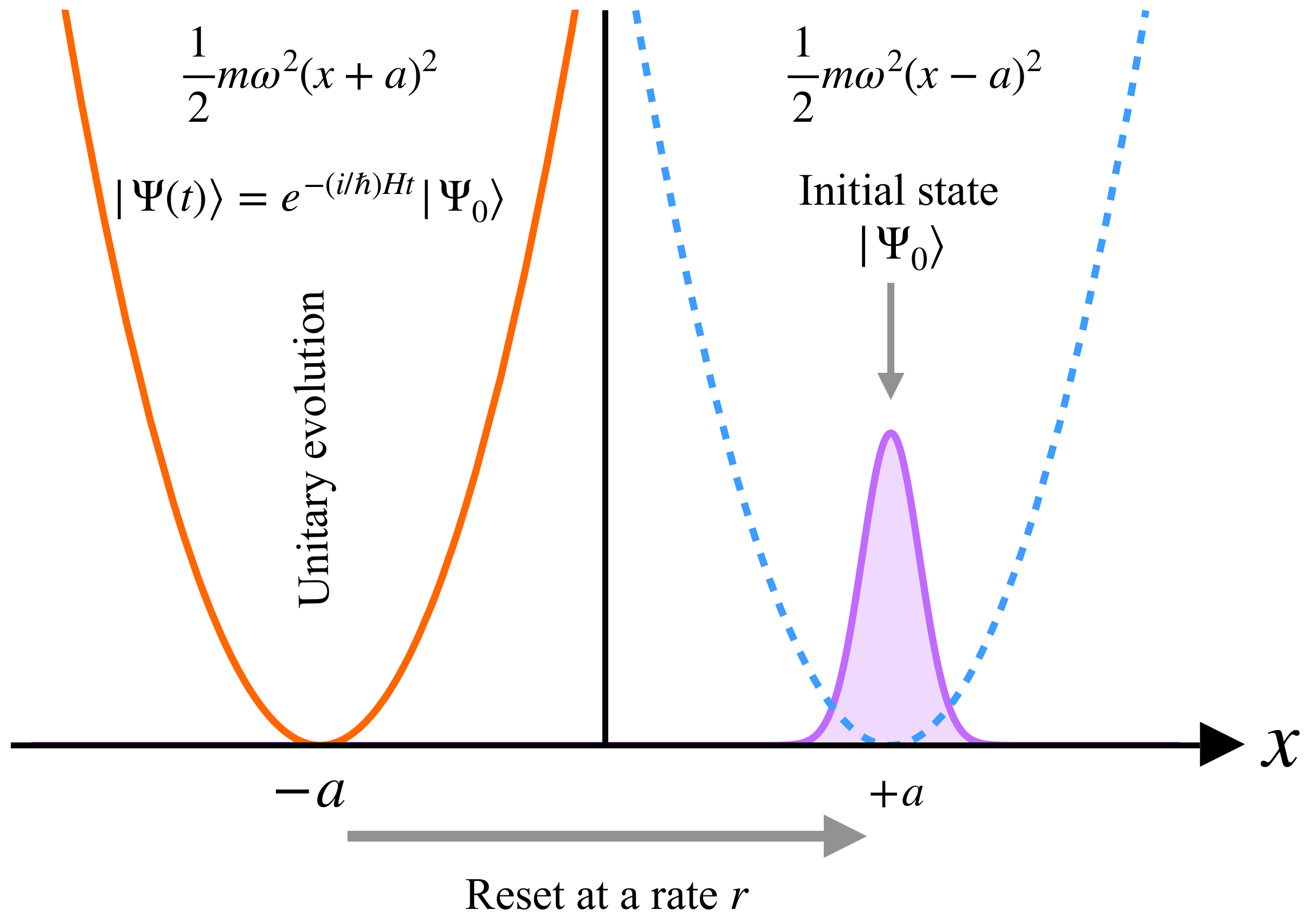}
    \caption{A schematic representation of the quantum resetting protocol. A system of $N$ noninteracting bosons is prepared in the ground state of a quantum harmonic oscillator of frequency $\omega$ and centered at $x=+a$. The system evolves unitarily with a harmonic oscillator Hamiltonian $H$ centered at $x=-a$, up to a random time drawn from an exponential distribution $r\, e^{-r\, \tau}$. After this random time, the trap center is moved instantaneously to $x=+a$, as shown by the arrow at the bottom, and the system is allowed to cool to the ground state -- this is the resetting to the initial state. The cycle is repeated again. This quantum resetting drives the system into a nonequilibrium steady state where the bosons get strongly correlated. } 
    \label{fig:schem}
\end{figure}

In our setup, the quantum resetting protocol is described as follows [see \fref{fig:schem}]. Initially, we prepare a system of $N$ noninteracting bosons in the ground state of a quantum harmonic oscillator (harmonic trap), of frequency $\omega$ and centered at $x=+a$. Such a system of noninteracting bosons may be realized experimentally in optical traps where the interactions between the bosons can be effectively tuned to zero via Feshbach resonance \cite{CCRCW2000,RZCMSIM07}.
We then instantaneously quench the trap center from $+a$ to $-a$ and let the system evolve unitarily up to a random Poissonian time $\tau$ distributed via $r\, e^{-r\tau}$. Following this, the trap center is instantaneously quenched back to $+a$ and the system is cooled to its initial ground state. This cooling can be achieved by coupling to a heat bath, and we do not make any measurements during this cooling period. This mimics instantaneous resetting as in the optical trap experiments in classical resetting~\cite{BBPMC20, FBPCM21}. This procedure is repeated, and eventually, the system reaches a NESS. Thus our protocol has three main parameters ($a$, $\omega$, $r$). We show that even though there is no direct interaction between the bosons, the repeated resetting of the trap center dynamically induces an effective all-to-all correlation between the bosons that persist even at long times. This becomes manifest in the fact that the quantum JPDF of $N$ bosons in the NESS is not factorizable. We considered this particular setup consisting of shifting the center of the oscillator from $a\to - a$ and resetting it back to $+a$ for two reasons. First, the initial state (centered around $x=+a$) is not an eigenstate of the Hamiltonian $H$. Secondly, the unitary evolution under $H$ preserves the Gaussian form of the initial state at all times, thereby making it amenable to analytical computations. 

The goal of this paper is to
show that the NESS reached under this quantum resetting protocol has the CIID structure, i.e., the quantum JPDF can be expressed as in \eref{eq:jointPDF}. This allows us to calculate several physical observables analytically in the large $N$ limit. These observables include the average particle density profile, two-point correlation functions, extreme and order statistics, 
distribution of the spacing between two adjacent particles (i.e., gap statistics), the statistics of the total number of particles in a given interval, i.e., the full counting statistics (FCS). Some of these observables are global, while the others are local. These observables have been studied extensively in one-dimensional classical correlated gases, e.g., in the 
eigenvalue statistics of a Gaussian random matrix~\cite{F10,M04} or its formulation as Dyson's log gas~\cite{MMSV14,LS17,MGreg14,MNSV09,AKD19}, Calogero model~\cite{AKD19}, Jellium model of one-dimensional one-component plasma~\cite{DKMSS17,DKMSS18,FMS21,FMS22}. More recently, such observables have also been studied in the Riesz gas~\cite{HLSS18,ADKKMMS19,KKK20, KKKMMS22,SKADKK22,KKKMMS24}. Observables such as the average
particle density and the statistics of the total number of particles in an interval can, in principle, be measured via absorption imaging techniques~\cite{K1999,SAFG11,HLSS2017,KA12}. On the other hand, local observables such as the order statistics, and in particular, the statistics of the rightmost particle can be potentially accessed by quantum gas microscopy~\cite{QGM1,QGM2,QGM3,BHMRT24}, where the resolution at the level of a single atom is feasible. Therefore, our proposed setup may be experimentally feasible, both in terms of designing the protocol as well as measuring the observables. \\

\noindent We briefly summarize our main findings: \\

\begin{enumerate}
    \item We show that under the quantum resetting protocol mentioned above, the system reaches a steady state where the quantum JPDF $P_r(x_1,x_2,x_3\cdots, x_N)$ exhibits a CIID structure \eref{eq:ren_NESS1U}, as in  \eref{eq:jointPDF}. We identify the random variable $u$ and its exact PDF $h(u)$ [see equation \eref{eq:hu3}]. 
\item We show that, in the steady state, the average density profile undergoes interesting shape transitions with respect to the resetting rate $r$ and the trap frequency $\omega$ for a fixed $a$ [see \fref{fig:dens}]. 

\item The two-point correlation function in the steady state is given by 
\begin{equation}
\label{eq:sum_corr}
C_{i,j} = \langle x_i x_j \rangle - \langle x_i  \rangle \langle  x_j \rangle = \frac{4 a^2 \left[5 (r/\omega) ^2+2\right]}{\left[(r/\omega) ^2+1\right]^2 \left[(r/\omega) ^2+4\right]} +\delta_{i,j}\frac{\hbar}{2\,m\, \omega}\, ,
\end{equation}
where $m$ is the mass and $\hbar$ is the Planck's constant. The function in \eref{eq:sum_corr} is positive, indicating an effective all-to-all attraction between bosons emerging due to simultaneous resetting. 

\item Strong correlations lead to extreme value and order statistics 
drastically different from that of independent and identically distributed (IID) and weakly correlated random variables [see  \fref{fig:os}]. 

\item The distribution of the number of bosons $N_L$ within the domain $[-L,L]$, known as the full counting statistics (FCS), shows rather interesting behaviour [see \fref{fig:fcs}]. For any finite $L$, there is a minimum fraction of  $N_L/N> 0 $ and a maximum fraction of  $N_L/N < 1$ beyond which the FCS vanishes in the thermodynamic limit $N\to \infty$. In other words, it has non-trivial finite support. In addition, remarkably, the FCS displays a discontinuity and an integrable divergence at an intermediate point inside the support [see \fref{fig:fcs}]. 

\end{enumerate}

The paper is organized as follows. In \sref{sec:mod_res}, we discuss the details of our model and the quantum resetting protocol. We obtain an interesting $N$-particle  JPDF that turns out to be of CIID form. In \sref{sec:obs}, we exploit the rich CIID structure to derive various observables. Perfect agreement is demonstrated between our analytical results and direct numerical simulations. We briefly describe the numerical simulation procedure used in the paper in \sref{sec:simulation}. In \sref{sec:conc}, we summarize our results along with an outlook. 

\section{The Model, the protocol, and the Quantum Resetting}
\label{sec:mod_res}

In this section, we first describe the steps involved in our protocol of quantum quench and resetting of $N$ noninteracting bosons in a harmonic trap. The protocol consists of the following steps:
\begin{enumerate}

\item We prepare the system of $N$ non-interacting bosons in the ground state  $|\Psi_0\rangle$ of a harmonic oscillator with a Hamiltonian given by 
\begin{equation}
H_0 = \sum_{j=1}^{N} \biggl[\frac{p_j^2}{2m} +\frac{1}{2}m\, \omega^2(x_j-a)^2 \biggr] \, ,
\label{eq:vx}
\end{equation}
where the harmonic potential is centered at $x=a>0$ and $p_i$ denotes the momentum if the $i$-th particle. 

\item We then instantaneously quench the system to a new Hamiltonian 
\begin{equation}
H = \sum_{j=1}^{N} \biggl[\frac{p_j^2}{2m} +\frac{1}{2}m\, \omega^2(x_j+a)^2 \biggr] \, ,
\label{eq:ham}
\end{equation}
where the potential is now centered at $x=-a$. 

\item Following this quench, the system evolves unitarily with the Hamiltonian $H$ up to a random time $\tau_1$ drawn from an exponential distribution $p_r(\tau_1) = r\,e^{-r\tau_1}$, where $r$ denotes the resetting rate. The state at the end of this step reads
\begin{equation}
    |\Psi(\tau_1)\rangle = e^{-\frac{i}{\hbar}H\tau_1} |\Psi_0\rangle\, .
\end{equation}

\item Following this unitary evolution up to $\tau_1$, the state $|\Psi(\tau_1)\rangle$ is reset to the same initial state $|\Psi_0\rangle$. In other words, the Hamiltonian is quenched back to $H_0$ and the system is allowed to relax to the ground state of $H_0$. We assume that this process $ |\Psi(\tau_1)\rangle \to |\Psi_0\rangle$ occurs instantaneously. 

\item We then repeat the steps (ii)--(iv). For each iteration, during the step (iii),  we choose the interval of unitary evolution $\tau$ independently from the exponential distribution 
$p_r(\tau) = r\,e^{-r\tau}$.

\end{enumerate}

The system evolving by this protocol approaches a NESS at long times. Our goal in this section is to characterize this NESS, i.e., to compute the quantum probability density associated with this gas of $N$ bosons. We show that the JPDF of the positions of $N$ particles can be computed exactly. We show explicitly that this JPDF does not factorize, indicating non-trivial correlations between the particles in the NESS. These correlations emerge purely from the dynamics since the particles have no direct interactions between them.  

Below, we describe in detail each of the steps above.

\subsection{Preparation of the initial state}
\label{sub:prep}

We prepare an initial state of $N$ noninteracting bosons in the ground state of the Hamiltonian $H_0$ given in \eref{eq:vx}.
The  many-body ground state wavefunction in the position basis is given by 
\begin{equation}
\langle x_1,x_2,\cdots, x_N|\Psi_0\rangle = \Psi_0(x_1,x_2,\cdots, x_N) = \prod_{j=1}^N \psi_0(x_j)\, 
\label{eq:Psi}
\end{equation}
where 
\begin{equation}
\psi_0(x) = A_0\,e^{-\frac{1}{2\sigma^2}(x-a)^2 }\,\quad \text{with}\quad \sigma = \sqrt{\frac{\hbar}{m \,\omega}}~~\text{and}~~A_0 = \Big( \frac{1}{\pi \sigma^2} \Big)^{1/4}\, . 
\label{eq:psi0}
\end{equation}
The subscript `0' in \eref{eq:Psi} and \eref{eq:psi0} stands for the `ground state'. The ground state in \eref{eq:Psi} is our initial state.

\subsection{Unitary evolution in the absence of resetting}
\label{sub:uni}

Starting from the initial state \eref{eq:Psi}, the system is evolved unitarily
by the Hamiltonian $H$ in \eref{eq:ham}. 
The subsequent time evolution of \eref{eq:Psi} according to \eref{eq:ham} takes the form
\begin{equation}
\Psi(x_1,x_2,\cdots, x_N,t) = \prod_{j=1}^N \psi (x_j,t)\, ,
\label{eq:Psit}
\end{equation}
where
\begin{equation}
\psi (x,t) = \int_{-\infty}^{\infty} K(x+a,y+a,t) \, \psi_0(y)\, dy \, .
\label{eq:Psit}
\end{equation}
The Mehler kernel $K(x,y,t) \equiv \langle x | e^{-\frac{i}{\hbar}H t}| y\rangle $ in \eref{eq:Psit} is given by~\cite{FH2010} 
\begin{equation}
\label{eq:k}
K(x,y,t) = \frac {1}{\sqrt{2\pi i \sigma^2 \sin (\omega t)}}\exp \left({\frac {i}{2\sigma^2\sin (\omega t)}}\left[(x^{2}+y^{2})\cos(\omega t)-2xy\right]\right)\, .
\end{equation}
Since the kernel and initial state are both of Gaussian form, the single-particle wavefunction $\psi (x,t)$ in \eref{eq:Psit} remains Gaussian at all times, i.e., 
\begin{equation}
\psi(x,t) = A(t) \, e^{-\frac{1}{2\sigma(t)^2}\big(x-\mu(t)\big)^2 }\, . 
\label{eq:psitg}
\end{equation}
Using \eref{eq:k} in \eref{eq:Psit}, we get 
\begin{equation}
\label{eq:mu_A_sigma}
\mu(t) = 2 a e^{-i\omega t} - a,\,~~\sigma(t) =\sigma,\,~~ A(t) = A_0 e^{-i\omega t/2} \exp\left[-\frac{a^2}{\sigma^2} \big(1-e^{-2 i\omega t}\big)\right]\, .
\end{equation}
As pointed out earlier, this preservation of the Gaussian structure during the unitary dynamics is our primary motivation for choosing this setup. 
The $N$-particle JPDF is given by 
\begin{equation}
P(x_1,x_2,\cdots, x_N,t)= |\Psi(x_1,x_2,\cdots, x_N,t)|^2 = \prod_{j=1}^N |\psi (x_j,t)|^2 = \, \prod_{j=1}^N p(x_j,t)\, , 
\label{eq:Psit2}
\end{equation}
where the single-particle PDF from \eqref{eq:psitg} and \eref{eq:mu_A_sigma} turns out to be 
\begin{equation}
p(x,t) = |\psi (x,t)|^2 = \frac{1}{\sqrt{\pi \sigma^2}}\,e^{-\frac{1}{\sigma^2}\big(x-\mu_R(t)\big)^2} \, , 
\label{eq:spdf}
\end{equation}
with $\mu_R (t)$ representing the real part of $\mu(t)$,
\begin{equation}
\label{eq:mur}
\mu_R (t) \equiv \mathrm{Re}[\mu(t)] = a \bigl(2 \cos(\omega t )-1 \bigr)\, . 
\end{equation}
Interestingly, the width $\sigma(t)=\sigma$ of the single-particle PDF is time independent whereas the location $\mu_R (t)$ where the Gaussian is centered, oscillates between $x = -3a$ and $x=+a$ with a time period $T = 2\pi/\omega$. Finally, the $N$-particle quantum JPDF takes the explicit factorized form 
\begin{equation}
P(x_1,x_2,\cdots, x_N,t)= \prod_{j=1}^N \, \frac{1}{\sqrt{\pi \sigma^2}}\,e^{-\frac{1}{\sigma^2}\big(x_j-\mu_R(t)\big)^2} \,,
\label{eq:Psit2_gauss}
\end{equation}
indicating the absence of any correlations among the bosons.

\subsection{Time evolution in the presence of quantum resetting}
\label{sub:res}

We now switch on the resetting with rate $r$. To see how the system evolves under the combined unitary 
`quantum' dynamics and the `classical' stochastic resetting dynamics, we start by recalling how the density matrix of a pure state evolves under resetting~\cite{MSM18}. Consider a system initially in a pure state $|\Psi_0\rangle$. Under the unitary evolution with a time-independent Hamiltonian $H$, the state evolves as $|\Psi(t)\rangle = e^{-(i/\hbar) H t} |\Psi_0\rangle$. Consequently, the density matrix $\varrho(t)=|\Psi(t)\rangle\langle\Psi(t)|$ evolves as 
\begin{equation}
    \varrho(t) = e^{-(i/\hbar) H t} \,\varrho(0) \,e^{(i/\hbar) H t}\quad\text{where}~~ \varrho(0)=|\Psi_0\rangle\langle\Psi_0|.
    \label{eq:formal}
\end{equation}
The quantum resetting is a mixture of classical stochastic and quantum unitary evolution defined as follows \cite{MSM18}. In a small time $dt$, the state of the system evolves as 
\begin{equation}
|\Psi (t+dt)\rangle = 
\begin{cases}
|\Psi_0\rangle &\quad\text{with  prob.}~~ r\,dt,\\[3mm] 
\bigl[1-(i/\hbar) H dt \bigr] \, |\Psi(t)\rangle &\quad\text{with  prob.}~~ 1-r\,dt\, ,
\end{cases}
\end{equation}
where $r$ represents the resetting rate. For $r=0$, one recovers the unitary evolution. Under this quantum resetting dynamics, one can show that the density matrix $\varrho_r (t)$ evolves as~\cite{MSM18} 
\begin{equation}
    \varrho_r (t) = e^{-r t} \, \varrho(t) + r \int_0^{t} d\tau e^{-r \tau} \varrho(\tau), 
    \label{eq:rho_ren}
\end{equation}
where $\varrho(t)$ is defined in \eqref{eq:formal} and the subscript `$r$' in $\varrho_r (t)$ stands for `resetting'. This result is easy to understand from the renewal nature of the underlying stochastic process. The first term corresponds to the case when there is no resetting event in the interval $[0,t]$ and the system evolves unitarily up to $t$. The second term corresponds to events where there are one or more resettings within $[0,t]$. In this latter case, it is enough to consider the epoch $t-\tau$ at which the last resetting event occurred before $t$. Then, during the interval $[t-\tau,t]$ the system evolves unitarily, explaining the presence of $\varrho(\tau)$ inside the integral of the second term in \eref{eq:rho_ren}. Finally, the probability that there is no resetting event in the interval $[t-\tau,t]$, preceded by a resetting event within an interval $d\tau$ at the beginning of this interval is simply $r \, d\tau \,e^{-r\tau}$. Finally, integrating over $\tau$ from $0$ to $t$, one gets the second term in \eref{eq:rho_ren}.

The quantum JPDF of $N$ particles with density matrix $\varrho_r(t)$ is given by the matrix element 
\begin{equation}
\label{eq:Pr_ren}
    P_r(x_1,x_2,\cdots, x_N,t) = \langle x_1, x_2, \dotsc, x_N|\varrho_r(t) |x_1, x_2, \dotsc, x_N\rangle.
\end{equation}  
Computing this matrix element from \eref{eq:rho_ren}, one gets
\begin{equation}
\mspace{-8mu}
P_r(x_1,x_2,\cdots, x_N,t)=  e^{-r t} P(x_1,x_2,\cdots, x_N,t) + r\int_0^t d\tau\, e^{-r \tau} P(x_1,x_2,\cdots, x_N,\tau)\, , 
\label{eq:ren}
\end{equation}
where $P(x_1,x_2,\cdots, x_N,t)$ is given by 
\begin{equation}
\mspace{-8mu}
    P(x_1,x_2,\cdots, x_N,t)= |\Psi(x_1,x_2,\cdots, x_N,t)|^2 = \langle x_1, x_2, \dotsc, x_N|\varrho(t) |x_1, x_2, \dotsc, x_N\rangle.
    \label{eq:dm}
\end{equation}
In our case, the quantum JPDF  in the absence of resetting $ P(x_1,x_2,\cdots, x_N,t)$, is given by \eref{eq:Psit2_gauss}.
 As $t\to\infty$, the first term in \eqref{eq:ren} drops out and one arrives at a nonequilibrium steady state given by 
\begin{equation}
P_r(x_1,x_2,\cdots, x_N)=   r\int_0^\infty d\tau\,e^{-r \tau} P(x_1,x_2,\cdots, x_N,\tau)\, . 
\label{eq:ren_NESS}
\end{equation}
Using \eref{eq:Psit2_gauss} in \eref{eq:ren_NESS}, we explictly get 
\begin{equation}
P_r(x_1,x_2,\cdots, x_N)=   r\int_0^\infty d\tau e^{-r \tau} \prod_{j=1}^N \, \frac{1}{\sqrt{\pi \sigma^2}}\,e^{-\frac{1}{\sigma^2}\big(x_j-\mu_R(\tau)\big)^2}\,  , 
\label{eq:ren_NESS1}
\end{equation}
where we recall that $\mu_R(\tau)=a (2 \cos(\omega \tau)-1)$ and $\sigma^2=\hbar/(m\omega)$. 
\Eref{eq:ren_NESS1} is the main result of this paper. It represents the steady state of the system undergoing resetting at a constant rate $r$ and   also has an alternative interpretation as follows: 
\begin{enumerate}
    \item We prepare the system in the ground state given by \eref{eq:Psi}. 
    \item We evolve it by the quantum Hamiltonian given in \eref{eq:ham} for a random time $\tau$ drawn from an exponential distribution $r\,e^{-r\tau}$.
    \item At the end of this time we make a measurement for the positions of particles.
    \item The process (i, ii, iii) above is repeated to reconstruct a JPDF of the position of the particles.
\end{enumerate}
This interpretation can be well suited for an experimental implementation as well as direct numerical simulations. It is evident from \eref{eq:ren_NESS1} that the JPDF in the presence of resetting does not have a trivial product form as was the case in \eref{eq:Psit2_gauss}, indicating strong correlations between the positions of the particles. As mentioned earlier, such dynamically emerging strong correlations between noninteracting particles with a CIID structure was recently investigated in several classical systems~\cite{BLMS_23,BLMS24,BMMS24,SM24}. 
However, such correlations in quantum gases have not been studied so far. 

As mentioned earlier, the observables of interest include the average particle density profile, two-point correlation functions, extreme and order statistics, gap statistics, and FCS. In order to compute them, it is useful to make a change of variable from $\tau$ to a new variable $u$ via the transformation 
\begin{equation}
\label{eq:umuR}
u = \mu_R(\tau) =  a \bigl(2 \cos(\omega \tau )-1 \bigr)\, \quad \text{where}~~u\in [-3 a,a]\, . 
\end{equation} 
As a result, \eref{eq:ren_NESS1} becomes 
\begin{equation}
P_r(x_1,x_2,\cdots, x_N)=   \int_{-3a}^{a} du\, h(u) \,\prod_{j=1}^N \, \frac{1}{\sqrt{\pi \sigma^2}}\,e^{-\frac{1}{\sigma^2}(x_j-u)^2}\, ,
\label{eq:ren_NESS1U}
\end{equation}
with
\begin{equation}
\label{eq:hu0}
    h(u) = \int_0^{\infty} \delta\left(u- a \left[2 \cos(\omega \tau )-1 \right]\right)\,r\,e^{-r \tau} \,d\tau = \sum_{n=1}^{\infty}\frac{r \, e^{-r \tau_n}}{ |du/d\tau|_{\tau_n}} \, ,
\end{equation}
where $\tau_n$ represents the $n$-th positive root of \eref{eq:umuR} for a given $u$. Evaluating $|du/d\tau|_{\tau_n}$ one gets
\begin{equation}
\label{eq:hu}
h(u) =  \sum_{n=1}^{\infty}\frac{r e^{-r\, \tau_n}}{2 a \omega|\sin(\omega \tau_n) |}\, . 
\end{equation}
The root $\tau_1$ in \eref{eq:hu} denotes the principal value of inverse cosine 
\begin{equation}
\label{eq:tau1}
\tau_1 = \frac{1}{\omega} \cos^{-1}\bigg[\frac{1}{2}\bigg(1+\frac{u}{a} \bigg)\bigg]\,\quad\text{where}~~\omega \, \tau_1 \in [0,\pi].
\end{equation}
The other roots of \eref{eq:umuR}, i.e., $\omega \, \tau_n \notin [0,\pi]$  are related to the principal root $\tau_1$ by 
\begin{align}
\label{eq:teven}
\tau_{2s} &= \frac{2 s\pi}{\omega} -\tau_1\quad   \text{where}~~\omega \, \tau_{2s} \in [(2s-1)\pi,2s \pi]\, ~~\quad \text{with}~~s=1,2,\cdots, \infty \\
\tau_{2s+1} &= \frac{2 s\pi}{\omega} +\tau_1\quad   \text{where}~~\omega \, \tau_{2s+1} \in [2s\pi,(2s+1) \pi]\,\quad \text{with}~~s=1,2,\cdots, \infty 
\label{eq:todd}
\end{align}
We note that $|\sin(\omega \tau_n)|$ that appears in \eref{eq:hu} is independent of index $n$ and is given by $|\sin(\omega \tau_n) | = \sqrt{1-\frac{1}{4}(1+u/a)^2}$. Therefore, \eref{eq:hu} becomes 
\begin{equation}
\label{eq:hu1}
h(u) = \frac{r/\omega}{ \sqrt{4a^2-(a+u)^2}} \sum_{n=1}^{\infty}e^{-r\, \tau_n}\, .
\end{equation}
The summation in \eref{eq:hu1} can be split into odd and even values of $n$. Subsequently using \eref{eq:teven} and \eref{eq:todd}, the summation can be performed explicitly and we get 
\begin{equation}
\label{eq:hu2}
h(u) = \frac{(r/\omega) \cosh\big[(r/\omega)(\pi-\omega\,\tau_1) \big]}{\sinh[\pi \,r/\omega] \sqrt{4a^2-(a+u)^2}}\, ,
\end{equation}
where $\tau_1$ is given in \eref{eq:tau1}. 
Note that in the limit $r\to 0$, the sum in \eref{eq:hu1} diverges and the amplitude $r/\omega\to 0$. However, their product tends to a non-zero value,
as seen from \eref{eq:hu2}, and is given by
\begin{equation}
\label{eq:smallr}
h(u) = \frac{1}{\pi \sqrt{4a^2-(a+u)^2}}\,\quad \text{for}~~ r\to 0\, . 
\end{equation}
Thus in this limit, using \eref{eq:ren_NESS1}, we get the exact result 
\begin{equation}
\lim_{r\to 0} \lim_{t\to\infty}  P_r(x_1,x_2,\cdots, x_N, t)=   \int_{-3a}^{a} du\, h(u) \,\prod_{j=1}^N \, \frac{1}{\sqrt{\pi \sigma^2}}\,e^{-\frac{1}{\sigma^2}(x_j-u)^2}\, ,
\label{eq:ren_NESS1Ux}
\end{equation}
with $h(u)$ given in \eref{eq:smallr}.
A physical explanation of this limiting $h(u)$ can be understood as follows. Since this result is obtained as $\lim_{r\to 0} \lim_{t\to \infty} P_r(x_1, x_2, \dotsc, x_N, t)$, one may wonder what happens if  instead one first takes the $r\to 0$ limit at fixed $t$ (i.e., no resetting) and then take  the $t\to\infty$ limit.  For $r=0$ at finite $t$, the distribution from \eqref{eq:Psit2_gauss} reads
\begin{equation}
    P_0(x_1, x_2,\dotsc, x_N, t) \equiv P(x_1, x_2,\dotsc, x_N, t)= \prod_{j=1}^N \frac{1}{\sqrt{\pi\sigma^2}} \, e^{-\frac{1}{\sigma^2}(x_j-a(2 \cos\omega t-1))^2}. 
    \label{rto0}
\end{equation}
This is clearly a time-periodic state with period $2\pi/\omega$. Clearly, the naive $t\to \infty$ limit does not exist and hence the two limits do not commute, i.e., 
\begin{equation}
    \lim_{r\to 0} \lim_{t\to\infty} P_r(x_1, x_2,\dotsc, x_N, t)\not=\lim_{t\to\infty} \lim_{r\to 0} P_r(x_1, x_2,\dotsc, x_N, t).
    \label{two-limits}
\end{equation}
To make the connection between the two ways of taking the limits, we note that the time-periodic state needs to be averaged over the period of oscillation $2\pi/\omega$, in order to make it time-independent, i.e., one should compare the left hand side of \eref{two-limits} with the time-averaged quantity
\begin{equation}
    \frac{\omega}{2\pi} \int_0^{2\pi/\omega} P_0(x_1, x_2, \dotsc, x_N, t)\, dt= \frac{\omega}{2\pi} \int_0^{2\pi/\omega} \prod_{j=1}^N \frac{1}{\sqrt{\pi\sigma^2}} \, e^{-\frac{1}{\sigma^2}(x_j-a(2 \cos\omega t-1))^2}\, dt.
    \label{average-tpss}
\end{equation}
By making the change of variable $u=a (2\cos(\omega t)-1)$, it is easy to see that the right hand side of \eref{average-tpss} is exactly identical to \eref{eq:ren_NESS1Ux} with $h(u)$ given in \eref{eq:smallr}. Hence, the we get the equality 
\begin{equation}
    \lim_{r\to 0} \lim_{t\to\infty} P_r(x_1, x_2,\dotsc, x_N, t) = \frac{\omega}{2\pi}\int_0^{2\pi/\omega} \lim_{r\to 0} P_r(x_1, x_2,\dotsc, x_N, \tau)\, d\tau.
    \label{eq:r0limit}
\end{equation}
The right hand side of \eref{eq:r0limit} can also be interpreted as evolving the free system ($r=0$) up to a random time $\tau$ drawn uniformly over the interval $\tau\in [0, 2\pi/\omega]$ and then averaging over $\tau$.

\begin{figure}
    \centering
    \includegraphics[width=0.7\linewidth]{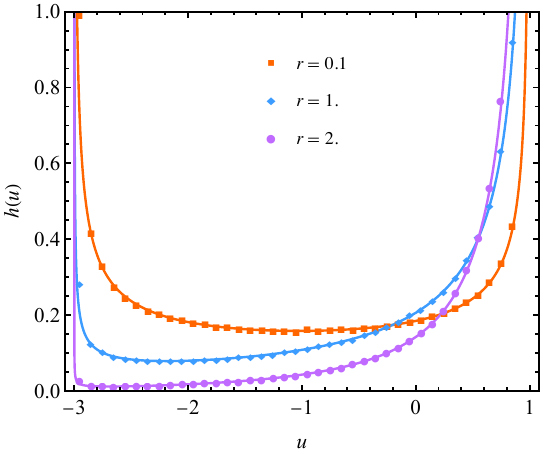}
    \caption{The solid lines plot the PDF $h(u)$ given in \eref{eq:hu3} as a function of $u$  for three different values of $r=0.1,\, 1.0, \,2.0$, keeping $\omega=1, a=1$. The points are obtained from numerical simulations with $64\times 10^4$ realizations.}
    \label{fig:hu}
\end{figure}

Reverting now to  general $r\ge 0$, we note that 
\eref{eq:hu2} further simplifies to 
\begin{equation}
\label{eq:hu3}
\mspace{-8mu}
h(u) = \frac{\tilde{r} }{\sinh(\pi \tilde{r}) } \frac{1}{\sqrt{4a^2-(a+u)^2}}\cosh\bigg[\tilde{r}\bigg(\pi-\cos^{-1}\Big(\frac{1}{2}(1+u/a)\Big)\bigg) \bigg]\, ~~\text{with}~~ \tilde{r}=\frac{r}{\omega}\, ,
\end{equation}
where $u \in [-3a,a]$.  In \fref{fig:hu}, we show the PDF $h(u)$ in \eref{eq:hu3} for three representative values of the resetting rate. Near the edges of the support $u \in [-3a,a]$, the function takes the form,
\begin{equation}
\label{eq:hlim}
h(u) =  \frac{\tilde{r} }{2 \sqrt{a}\,\sinh(\pi \tilde{r}) }  \times \begin{cases}
\displaystyle
\frac{1}{\sqrt{3 a+u}} + O(\sqrt{3 a+u}) &\quad \text{as}~~u\rightarrow -3a \, ,\\[6mm]
\displaystyle
\frac{\cosh (\pi  \tilde{r})}{ \sqrt{a-u}}
+ O(\sqrt{a-u}) &\quad \text{as}~~u\rightarrow a\, . 
\end{cases}
\end{equation}
Here, we remark that interestingly the JPDF structure in \eref{eq:ren_NESS1U} was recently obtained for a classical gas of non-interacting particles in a harmonic potential whose center is driven by an independent stochastic process \cite{SM24}. It is however, not at all evident whether $h(u)$ derived in this work [see equation \eref{eq:hu3}] can be obtained from a noninteracting classical gas studied in Ref.~\cite{SM24}. Nevertheless, the similarity in the mathematical structure in the JPDF enables us to adapt many of the approaches in Ref.~\cite{SM24} to compute some of the observables mentioned above. Starting with \eref{eq:ren_NESS1U} and \eref{eq:hu3}, we evaluate several observables in Sec.~\ref{sec:obs}. 

\section{Observables}
\label{sec:obs}
In this section, we discuss various observables for the correlated bosonic gas.  

\subsection{Average density profile}
\label{subsec:dp}
\begin{figure}
    \centering
    \includegraphics[width=0.5\linewidth]{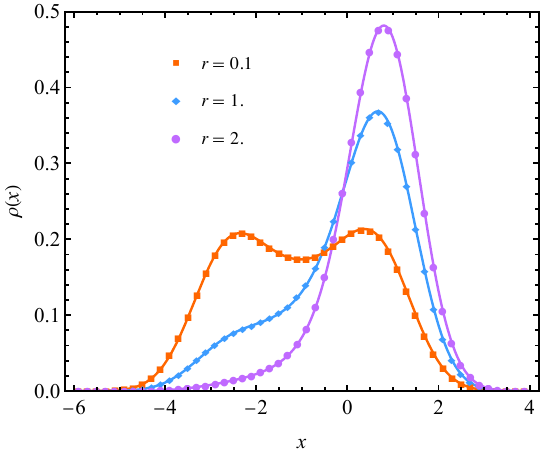}~~\includegraphics[width=0.5\linewidth]{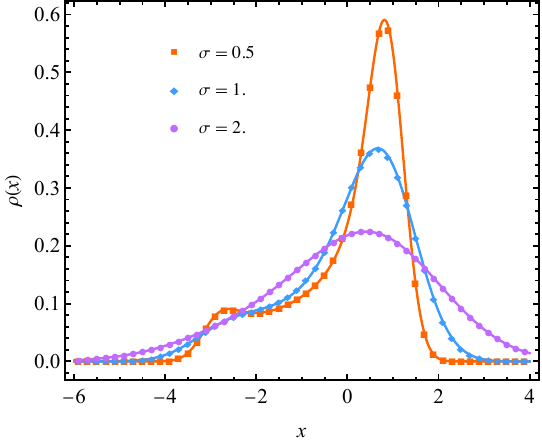}
    \caption{(Left) The spatial density profile $\rho(x)$ vs. $x$ in the steady state with $\sigma=1$ for $r = 0.1,\, 1.0,\, 2.0$.  (Right) The spatial density profile with $r=1$ for $\sigma = 0.1,\, 1.0,\, 2.0$. The solid lines plot equation~\eref{density1} and the points are obtained from numerical simulations with $64\times 10^4$ realizations and $N=10^4$.}
    \label{fig:dens}
\end{figure}

A natural observable that enables one to characterize a cloud of gas is the average density profile,
\begin{equation}
\label{eq:dens}
\rho(x) = \left\langle \frac{1}{N} \sum_{i=1}^N \delta(x-x_i) \right\rangle.
\end{equation}
where the average $\langle \cdots \rangle$ is with respect to the JPDF given in \eref{eq:ren_NESS1U}. Evaluating \eref{eq:dens}, we get 
\begin{equation}
\rho(x) = 
\frac{1}{\sqrt{\pi \sigma^2}} \, 
\int_{-3 a}^{a} du\, h(u)\,\exp\left(-\frac{(x-u)^2}{\sigma^2}\right)
\, ,
\label{density1}
\end{equation}
where $h(u)$ is given in \eref{eq:hu3}.
Furthermore, the density in \eref{density1} can be expressed as a scaling function of $x/a$ together with two independent dimensionless parameters $r/\omega$ and $\sigma/a$. This is essentially equivalent to setting $\omega=1$ and $a=1$.  In \fref{fig:dens}, we plot the density in \eref{density1} for various values of $r$ for a fixed $\sigma$ and various values of $\sigma$ for a fixed $r$. We find interesting shape transitions with respect to both $r$ and $\sigma$. For example, by fixing $\sigma$ and varying $r$, on finds that there is a critical resetting rate $r^*$ across which the density profile exhibits a bimodal to a unimodal transition. To locate this transition, we set $\rho'(x^*) = 0$ and $\rho''(x^*) = 0$ where $x^*$ is the inflection point and $\rho(x)$ in \eref{density1}. These two equations determine both $x^*$ and $r^*$ uniquely. Likewise, for a fixed $r$, there is a critical $\sigma^*$ where the density profile undergoes a similar transition.

\begin{figure}
    \centering
\includegraphics[width=0.37\linewidth]{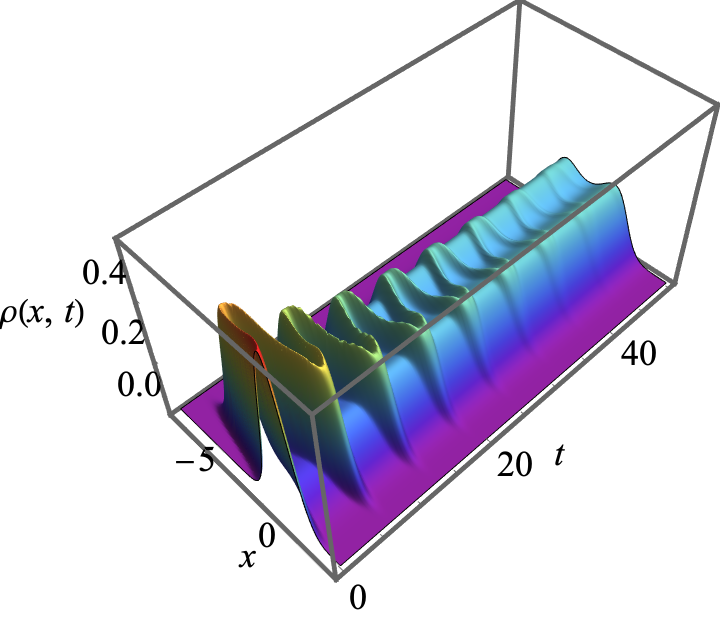}
~~\includegraphics[width=0.28\linewidth]{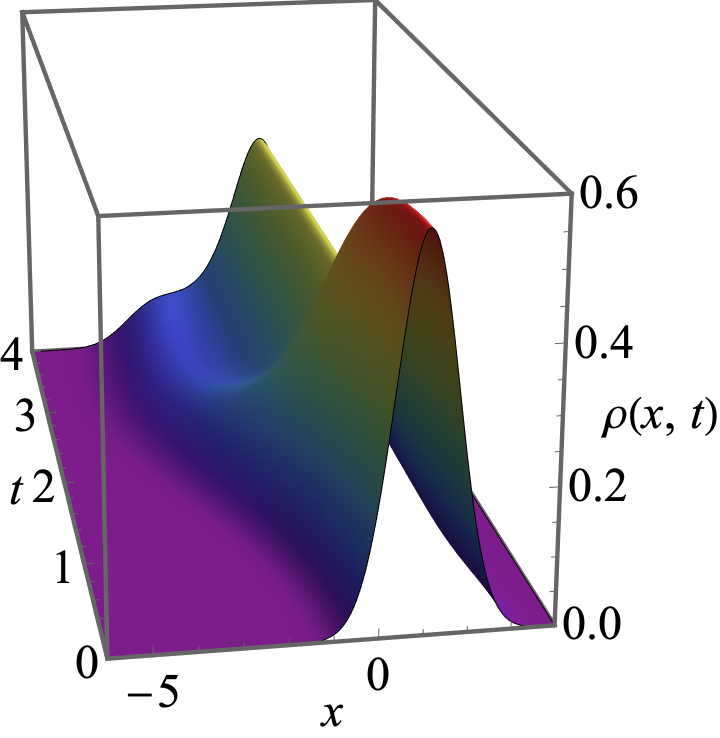}~~~~\includegraphics[width=0.28\linewidth]{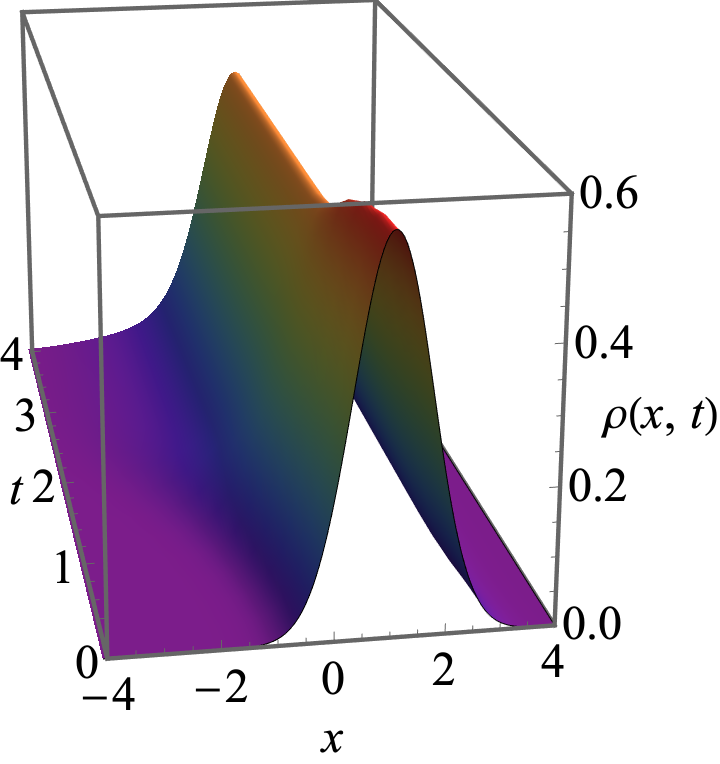}
    \caption{Space-time plots of the average density profile $\rho(x,t)$ vs. $x$ with time $t$ as the other axis, for   $r=0.1$ (left),    $r=1.0$ (middle), and  $r=2.0$ (right). The figure shows that the density profile in the presence of resetting given by \eref{eq:rhoxrt} oscillates in time, and the oscillations eventually diminish as time progresses. Eventually, the profile reaches a corresponding steady state shown in \fref{fig:dens}. We set $a=1$ and $\sigma=1$. The orientations of the figures are chosen to make the features best visible.}
    \label{fig:time}
\end{figure}

To understand this shape transition, it is useful to look at the two extreme limits $r\to 0$ and $r \to \infty$. The limit $r\to 0$ has already been discussed in detail in the previous section.  In this limit, substituting the expression of $h(u)$ from \eref{eq:smallr} into \eref{density1}, we get the stationary density as
\begin{equation}
\rho(x) = 
\frac{1}{\sqrt{\pi \sigma^2}} \, 
\int_{-3 a}^{a} du\,  \frac{1}{\pi \sqrt{4a^2-(a+u)^2}}\,\exp\left(-\frac{(x-u)^2}{\sigma^2}\right)
, 
\label{density1sr}
\end{equation}
which is a bimodal distribution for $\sigma/a < 1.591\cdots$ and unimodal otherwise. On the other hand, in the $r\to \infty$, we observe the system mostly in the initial state, which is centered around $x=a$, giving a unimodal structure to the distribution. Indeed, in the $r\to \infty$ limit, $h(u)$ in \eref{eq:hu3} approaches a Dirac delta function $\delta(u-a)$ resulting in $\rho(x) = 1/(\sigma\sqrt{\pi})\,e^{-(x-a)^2/\sigma^2}$ for any $\sigma$. Therefore, as we interpolate from small to large $r$, the density profile undergoes a bimodal to a unimodal shape transition. \\

\noindent
\textit{Approach to the steady state:} So far, we have discussed the steady state properties of the system. However, our approach can be easily adapted to also study how the system relaxes to the stationary state at long times. For instance, the average density profile at any time $t$ is given by 
\begin{equation}
\label{eq:rhoxrt}
\rho(x,t) = e^{-r t}\, p(x,t) +r\int_0^{t} d\tau \, e^{-r\tau} \, p(x,t)\, ,
\end{equation}
where $p(x,t)$ is the single-particle PDF in the absence of resetting given in \eref{eq:spdf} with \eref{eq:mur}.
Unfortunately, the integral in \eref{eq:rhoxrt} can not be carried out explicitly for finite $t$. However, one can easily plot $\rho(x,t)$ as a function of $x$, for fixed $t$, to see how the average density profile evolves with time and approaches the stationary profile as $t \to \infty$. This is shown in \fref{fig:time} for three values of the resetting rate $r$. 

\subsection{Correlation functions}

In this subsection, we will discuss connected two-point correlation function between the positions of the particles. It is defined in the stationary state as
\begin{equation}
    C_{i,j} = \langle x_i x_j \rangle - \langle x_i  \rangle \langle  x_j \rangle,
    \label{eq:correlation}
\end{equation}
where $\langle \cdots \rangle$ is over the stationary measure \eref{eq:ren_NESS1U}. Following Ref.~\cite{SM24}, we get 
\begin{equation}
    C_{i, j} =
     \mathrm{Var}(u) + \delta_{i,j}\,  \frac{\sigma^2}{2} \, , 
    \label{Cij}
\end{equation}
where $ \mathrm{Var}(u)$ is the variance of $u$ with respect to the stationary PDF $h(u)$, i.e., 
\begin{equation}
    \mathrm{Var}(u) = \langle u^2 \rangle - \langle u \rangle^2 = 
    \int_{-\infty}^\infty u^2\, h(u)\, du - \left[ \int_{-\infty}^\infty u\, h(u)\, du \right]^2.
\end{equation}
Using $h(u)$ from \eref{eq:hu3}, we get 
\begin{equation}
\label{eq:var}
\mathrm{Var}(u) = \frac{4 a^2 \left(\tilde{r}^2+2\right)}{\left(\tilde{r}^2+1\right)^2 \left(\tilde{r}^2+4\right)}\,\quad\text{with}~~ \tilde{r}=\frac{r}{\omega} .
\end{equation}
Interestingly $\mathrm{Var}(u)$ in \eref{eq:var} is non-monotonic in $\tilde{r}$. It initially increases quadratically from $2a^2$, reaches a maximum at $\tilde{r} = 0.2734\dots$, and then decreases monotonically with increasing $\tilde{r}$, eventually as $\tilde{r}^{-4}$ for large $\tilde{r}$.

\begin{figure}
    \centering
    \includegraphics[width=0.5\linewidth]{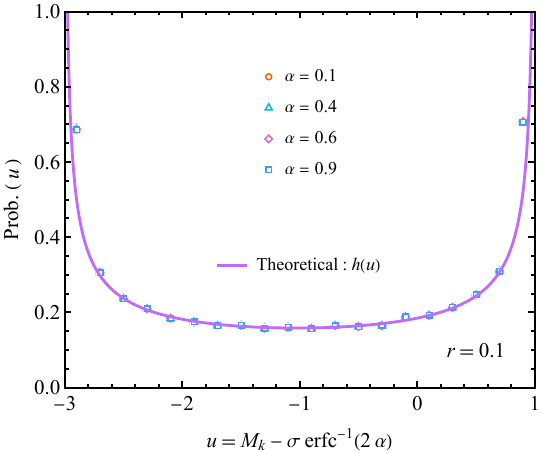}~~\includegraphics[width=0.5\linewidth]{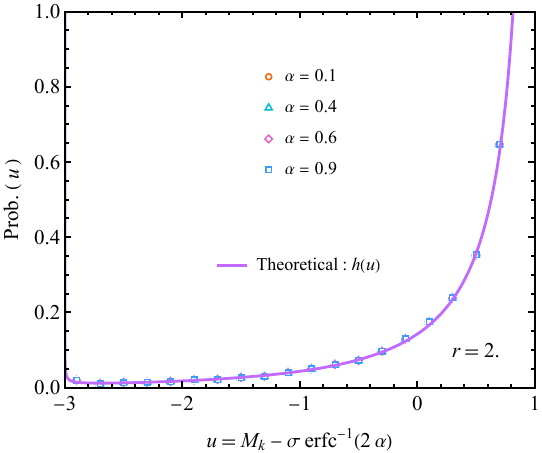}
    \caption{Order statistics, $\mathrm{Prob.}[M_k-\sigma\mathrm{erfc}^{-1}(2\alpha)=u]$ vs. $u$,  for $r=0.1$ (left) and $r=2.0$ (right) for different values of $\alpha=k/N$ ranging from the right edge to the left edge via the bulk. The solid (theoretical) lines plot equation~\eref{Mk-distribution} where $h(u)$ is given by \eqref{eq:hu3}, and the points are obtained from numerical simulations with $7\times 10^5$ realizations and $N=10^6$. We set $\omega=1$, $\sigma=1$ and $a=1$.}
    \label{fig:os}
\end{figure}

\subsection{Order and gap statistics}
\label{subsec:os}

While the average density profile discussed in~\sref{subsec:dp} gives the macroscopic spatial profile of the particles, it does not give any information on the typical microscopic spatial structure of individial particles. These microscopic fluctuations in the position can be probed by other observables as discussed below. In fact, since the sytem is inhomogenous due to the presence of the harmonic trap, one may expect that the microscopic fluctuations near the trap center (denoted by ``bulk") may actually differ from those that are far away from the center (loosely referred to as ``edge").

To unravel these features, we first investigate the order statistics, i.e., the fluctuations of the ordered position $M_k$ of the $k-$th particle from the right. In other words, particle $M_1 = \max\{x_1,x_2,\cdots, x_N\}$ denotes the position of the rightmost particle. By choosing $k\sim O(1)$, one probes the behaviour of the particles near the ``edge", while by choosing $k\sim O(N)$, one probes the ``bulk". Subsequently, the PDF of the $M_k$ is given by
\begin{equation}
    \mathrm{Prob.}(M_k=w) =
    \int_{-3a}^{a} du\, h(u)\,
    \mathrm{Prob.} (M_k(u)=w),
    \label{eq:maximum-dist}
\end{equation}
where $M_k(u)$ is the position of the $k-$th maxima for a set of $N$ IID random variables drawn from a Gaussian distribution with mean $u$ and variance $\sigma^2/2$. 
In addition to the order statistics, we present results for the statistics of the gap between $k-$th and $(k+1)-$th particle. Note that $M_k$ is a continuous variable and therefore, with a slight abuse of notation, we denote the PDF of $M_k$ by $\mathrm{Prob.}(M_k=w)$ such that $\int dw\, \mathrm{Prob.}(M_k=w) = 1$. Throughout this paper, we use this notation $\mathrm{Prob.}$ to denote the PDF of a continuous random variable. 

\subsubsection{Distribution of the maximum (rightmost particle):}
\label{subsub:max}

It is well known from the theory of extreme value statistics~\cite{MPS20, MS24, BLMS_23,SM24, SS19, MRS08} that the distribution of the maximum of a set of Gaussian IID random variables, around its typical value $u+\sigma  \,\sqrt{\ln N}$, is given by the Gumbel distribution whose width goes to zero as $1/\sqrt{\ln N}$. Therefore in the large-$N$ limit,  
\begin{math}
  \mathrm{Prob.}(M_1(u)=w) \simeq  \delta(w-u - \sigma^2\, \sqrt{\ln N} ). 
\end{math}
As a consequence, from \eref{eq:maximum-dist} with $k=1$, we get 
\begin{equation}
    \mathrm{Prob.}(M_1=w)\simeq  h\left(w-\sigma\,\sqrt{\ln N}  \right)\, ,
    \label{eq:maxCIID}
\end{equation}
where $h(u)$ in given in \Eref{eq:hu3}.

\subsubsection{Distribution of the $k$-th maximum}
\label{subsub:kth}
Again from the theory of order statistics, the typical value of the $k$-th maximum of a set of IID Gaussian random variables is given by~\cite{MPS20, MS24, BLMS_23,SM24, MRS08}  $ w^*(u) = u +\sigma\, \mathrm{erfc}^{-1}(2\alpha)$, where $\alpha=k/N$ and $\mathrm{erfc}^{-1}$ is the inverse complementary error function, i.e., $\mathrm{erfc}[\mathrm{erfc}^{-1} (z)]=z$. Furthermore, the fluctuation around the typical value $w^*(u)$ goes to zero as $N\to\infty$. As a consequence, for large $N$, we have $\mathrm{Prob.}(M_k(u)=w) \simeq\delta(w-w^*(u)) $. 
Finally from \eref{eq:maximum-dist}, we get 
\begin{equation}
    \mathrm{Prob.}(M_k=w) \simeq h\left(w-\sigma\, \mathrm{erfc}^{-1}(2\alpha)\right).
    \label{Mk-distribution}
\end{equation}
In \fref{fig:os}, we compare our analytical predictions of the order statistics given by \eqref{Mk-distribution} with numerical simulations for various values of $\alpha$ and find excellent agreement. 

\begin{figure}
    \centering
    \includegraphics[width=0.7\linewidth]{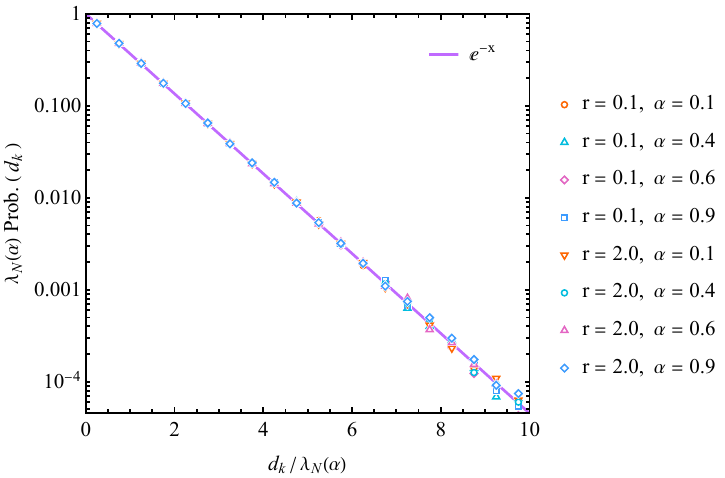}
    \caption{Scaling data collapse for the distribution $\mathrm{Prob.}(d_k)$ of the $k$-th gap $d_k$ as in~\eref{eq:gap-dist} for different values of $\alpha=k/N$, $\sigma$, and $r$. The characterstic gap size $\lambda_N(\alpha)$ is given in \eqref{lambda-alpha}.  The points are obtained from numerical simulation with $7\times 10^5$ realizations and $N=10^6$. We set $\omega=1$ and $a=1$. The solid line plots the exponential function $e^{-x}$.}
    \label{fig:gap}
\end{figure}

\subsubsection{Gap statistics:}
\label{subsub:gap}
We now discuss the statistics of the gap $d_k$ between the $k$-th and $(k+1)$-th particle. 
Let $d_k(u)=M_k(u)-M_{k+1}(u)$ represents the gap between the position of the $k$-th particle for a given $u$. Therefore, averaging over $u$, we get
\begin{equation}
    \mathrm{Prob.}(d_k=g) = \int_{-3a}^{a} du\, h(u)\,
    \mathrm{Prob.} \bigl(M_k(u)-M_{k+1}(u)=g\bigr).
    \label{dk}
\end{equation}
As described in Ref.~\cite{SM24}, the statistics of the gap is fully independent of the PDF $h(u)$ given in \eqref{eq:hu3}. This is because the gap between particle positions does not depend on the shift parameter $u$ which is common to all the $x_i$'s in \eref{eq:ren_NESS1U}. As a result, we get the same distribution as given in Ref.~\cite{SM24} which reads
\begin{equation}
    \mathrm{Prob.}(d_k=g) 
   \simeq \frac{1}{\lambda_N(\alpha)} \exp\left(-\frac{g}{\lambda_N(\alpha)}\right),
   \label{eq:gap-dist}
   \end{equation}
where the characteristic gap size $\lambda_N(\alpha)$ is given by
\begin{equation}
\lambda_N(\alpha) = \left[ 
   \frac{N}{\sigma\sqrt{\pi }} \exp\bigl(- [\mathrm{erfc}^{-1}(2\alpha)]^2\bigr)
   \right]^{-1}.
   \label{lambda-alpha}
\end{equation}

In \fref{fig:gap}, we compare the theoretical prediction given in \eref{eq:gap-dist} with numerical simulations and find excellent agreement for various values $\sigma$. Furthermore, we demonstrate that it is indeed independent of resetting rate $r$.

\subsection{Full Counting Statistics}
\label{sub:fcs}

This section is devoted to the study of the full counting statistics (FCS) that describes the distribution of the number of particles $N_L$ in a given region $[-L,L]$.  Evidently, $N_L$ is a random variable, and it will be interesting to study the probability distribution $P(N_L,N)$, for the correlated gas of $N$ bosons described by the JPDF in \eref{eq:ren_NESS1U}. We show below that $P(N_L,N)$ exhibits very interesting behaviours. First note that the $\langle N_L \rangle = N\times \int_{-L}^{+L} \rho(x)\, dx$ where $\rho(x)$ is the average density given in \eqref{density1}. Thus, the typical scale of the random variable $N_L$ is clearly set by $N$. Hence, for large $N$ and large $N_L$, one expects a natural scaling form 
\begin{equation}
  P(N_L,N) \simeq \frac{1}{N} H\left(\frac{N_L}{N}\right)\, \,
  \label{FCS-scaling0}
\end{equation}
where the fraction $N_L/N = \kappa$ lies in $\kappa \in [0,1]$ and $H(\kappa)$ is the scaling function. The factor $1/N$ multiplying $H(\kappa)$ in \eqref{FCS-scaling0} ensures the normalization $\int_0^1 H(\kappa) d\kappa =1$. We will compute the scaling function $H(\kappa)$ explicitly later, but let us already mention two rather interesting and surprising features of $H(\kappa)$ below:
 
\begin{itemize}

\item We show that the scaling function $H(\kappa)$ is supported over a finite interval $\kappa \in [\kappa_{\rm min}, \kappa_{\rm max}]$ that is different from $[0,1]$, i.e.,  $\kappa_{\rm min} > 0$ and $\kappa_{\rm max}<1$. Thus the fraction of particles that get accommodated in the region $[-L,L]$ cannot be less than $\kappa_{\rm min}$ and also cannot be bigger than $\kappa_{\rm max}$. This is rooted in the fact that strong correlations are present in the gas. While for finite-$N$, the full range of $\kappa \in [0,1]$ is allowed, the support $\kappa \in [\kappa_{\rm min}, \kappa_{\rm max}]$ becomes smaller only in the thermodynamic limit $N\to \infty$. A similar fact was noticed earlier in the FCS of classical systems~\cite{BMMS24, SM24}. 

\item A second dramatic feature of the FCS in our model is that there is an intermediate value $\kappa_{\rm min } < \kappa^{*} < \kappa_{\rm max}$ such that the scaling function $H(\kappa)$ displays different behaviours in the two regimes: (I) $\kappa_{\rm min} \leq \kappa < \kappa^* $ and (II) $\kappa^* \leq \kappa \leq \kappa_{\rm max} $. Moreover, the scaling function exhibits a discontinuity at $\kappa = \kappa^{*}$ [see \fref{fig:fcs}]. The scaling function  $H(\kappa)$ approaches a constant as $\kappa$ approaches $\kappa^*$ from below and has an integrable divergence as $\kappa$ approaches $\kappa^*$ from above. Below, we will discuss the two regimes separately. 
\end{itemize}

\begin{figure}[t!]
    \centering
    \includegraphics[width=0.7\linewidth]{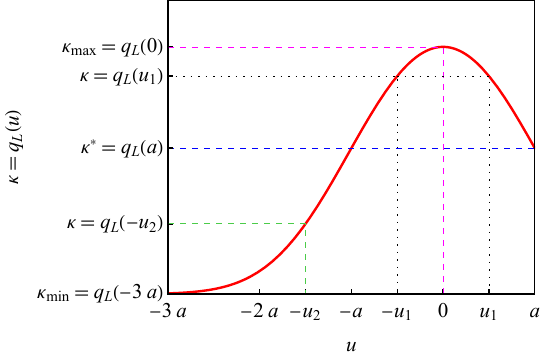}
    \caption{Plot of $\kappa = q_L(u)$ as a function of $u$, where $q_L(u)$ is given in \eref{eq:qL}. As the range of $u$ in the integral given in \eref{FCS-scaling} lies in $[-3a, a]$, the lower support of  $\kappa$ is given by $\kappa_{\rm min} = q_L(-3 a)$ and is strictly positive. Similarly, the upper support of $\kappa$ is given by $\kappa_{\rm max} = q_L(0)$ and is strictly less than unity for any $L < \infty$. It is important to note that for $\kappa_{\rm min}\le  \kappa <\kappa^*=q_L(-a)$, there is a unique $u$ for a given $\kappa$ (represented by green dashed line). On the contrary, for $\kappa^* \le \kappa \le  \kappa_{\rm max}$, two values of $u$ contribute to the integral in \eref{FCS-scaling} for a given $\kappa$ (represented by black dotted line).} 
    \label{fig:qlu}
\end{figure}

Using the CIID structure of the JPDF in \eref{eq:ren_NESS1U} and \eref{eq:hu3}, one can express the FCS as 
\begin{equation}
    P(N_L,N)= \int_{-3a}^{a} du\, h(u)\, P(N_L,N|u)\, , 
    \label{eq:PNL}
\end{equation}
where $P(N_L,N|u)$ is the FCS for a set of $N$ IID random variables drawn from a Gaussian distribution with mean $u$ and variance $\sigma^2/2$. For this set of IID Gaussian random variables, the probability of finding each particle within the interval $[-L,L]$ is evidently, 
\begin{equation}
    q_L(u)=\int_{-L}^L \frac{1}{\sqrt{\pi \sigma^2}}\,e^{-\frac{1}{\sigma^2}(x_j-u)^2}\, dx =\frac{1}{2} \left(\mathrm{erf}\left[\frac{  L-u}{\sigma }\right]+\mathrm{erf}\left[\frac{ L+u}{\sigma }\right]\right),
    \label{eq:qL}
\end{equation}
where $\text{erf}(x)$ is the error function. 
Therefore, $P(N_L,N|u)$ is simply given by the binomial distribution,
\begin{equation}
    P(N_L,N|u) = \binom{N}{N_L}
    [q_L(u)]^{N_L} [1-q_L(u)]^{N-N_L}.
    \label{FCS-dist-cond}
\end{equation}
Before proceeding further, let us recall $\kappa = N_L/N$.  It is straightforward to show from the Binomial distribution in \eqref{FCS-dist-cond} that the distribution of $\kappa$ converges to a Gaussian distribution centered around $q_L(u)$ with a width proportional to $1/\sqrt{N}$. Hence in the large-$N$ limit, equation~\eref{FCS-dist-cond} becomes
\begin{equation}
   P(N_L=\kappa N,N|u)\to\frac{1}{N}\,\delta\left(
    \kappa -q_L(u)  \right) .  
\end{equation}
Consequently,
\begin{equation}
  P(N_L,N) \simeq \frac{1}{N} H\left(\frac{N_L}{N}\right)\quad\text{with}~~  H(\kappa)=   \int_{-3a}^{a} du\, h(u)\, 
  \delta[\kappa - q_L(u)],
  \label{FCS-scaling}
\end{equation}
where $q_L(u)$ and $h(u)$ are given in \eref{eq:qL} and \eref{eq:hu3} respectively. It is evident from \eref{eq:qL} that $q_L(u) \to 1$ as $L\to \infty$. As a result from \eqref{FCS-scaling}, we get $H(\kappa) \to \delta(\kappa-1)$. This is consistent with the fact that for an infinite domain the probability of finding $N_L=N$ particles is obviously unity.

Below, we will see that all the interesting features of the scaling function $H(\kappa)$, mentioned in the beginning of this section, emerge from performing the integral in \eref{FCS-scaling} for any finite $L$. 
In order to carry out the integral over delta function in \eref{FCS-scaling}, it is useful to plot $ q_L(u)=\kappa$ in \eref{eq:qL} as a function of $u$. This is shown in \fref{fig:qlu} where the allowed range of $u$ is $[-3a, a]$. When $u=-3a$, i.e., at the lowest allowed point, the corresponding value of $\kappa$ is given by 
\begin{equation}
\label{eq:kmin}
\kappa_{\rm min} = q_L(-3a)\, . 
\end{equation}
Similarly, the maximum allowed value of $\kappa$ for $u\in [-3a, a]$ occurs at the maximum of the function $q_L(u)$, i.e.,  at $u=0$, and is given by  
\begin{equation}
\label{eq:kmax}
\kappa_{\rm max} = q_L(0)\, . 
\end{equation}
We notice from \fref{fig:qlu} that the function $q_L(u)$ is non-monotonic in the range $u\in [-3a, a]$. As a result, the inverse function $u = q_L^{-1}(\kappa)$ has a single root for $\kappa_{\rm min} \leq \kappa < \kappa^*$ (regime I), while it has two roots for $\kappa^* \leq \kappa \leq \kappa_{\rm max}$ (regime II), where $\kappa^*$ is given by 
\begin{equation}
\label{eq:ks}
\kappa^{*} = q_L(-a) = q_L(a). 
\end{equation}
Therefore, we analyse $H(\kappa)$ in \eref{FCS-scaling} separately in the two regimes I and II. 

\subsubsection{Regime I: $\kappa_{\rm min} \leq \kappa <\kappa^*$: }
\label{subsubk1}

Given that there is a unique value of $u\in [-3a, -a)$ for a given $\kappa$ in this regime, we get 
\begin{equation}
\label{eq:bks}
H(\kappa)=   \int_{-3a}^{-a} du\, h(u)\, 
  \delta[\kappa - q_L(u)]= \int_{-3a}^{-a} du\, h(u) \frac{\delta(u-q_L^{-1}(k))}{|q_L'(u)|} = \frac{h\bigl(q_L^{-1}(k)\bigr)}{q_L'\bigl(q_L^{-1}(k)\bigr)}\, , 
\end{equation}
where we have removed the modulus in the last step of \eref{eq:bks} since $q_L'(u) > 0$ for $u\in [-3a,a)$.
Using the expression of $q_L(u)$ in  \eref{eq:qL},  equation~\eref{eq:bks} further simplifies to 
\begin{equation}
\label{eq:bks1}
H(\kappa)=  \frac{   \sigma\sqrt{\pi}\,h\bigl(q_L^{-1}(k)\bigr) }{\exp\left({-\frac{\bigl(L+q_L^{-1}(k)\bigr)^2}{\sigma ^2}}\right)- \exp\left({-\frac{\bigl(L-q_L^{-1}(k)\bigr)^2}{\sigma ^2}}\right)}\, \quad \text{where}~~\kappa_{\rm min}\leq \kappa <\kappa^*,
\end{equation}
with $q_L^{-1}(k)$ being the inverse of the function $q_L(u)$ given in \eref{eq:qL}, in the range $u\in [-3a ,-a)$. 

We now analyze $H(\kappa)$ given in \eref{eq:bks1} near the two edges $\kappa_{\rm min}$ and $\kappa^{*}$ . We recall that as  $\kappa \to \kappa_{\rm min}$, i.e., $q_L^{-1}(\kappa)\equiv u \to -3a$, the limiting behavior of the function $h(u)$ is given in \eref{eq:hlim}, i.e., $h(u)$ diverges as $1/\sqrt{u+3a}$ as $u\to -3a$. Expanding $q_L(u)$ in Taylor series near $u=-3a$, we get 
\begin{equation}
\label{eq:up3a}
    u+3a = \frac{\kappa-\kappa_{\rm min}}{q_L'(-3a)}\quad\text{where}~~ q_L(u)=
    \kappa~~\text{and}~~ q_L(-3a) = \kappa_{\rm min}.
\end{equation}
Therefore using \eref{eq:up3a} in \eref{eq:hlim} we get  
\begin{equation}
\label{eq:hqi}
    h\bigl(q_L^{-1}(k)\bigr) \simeq  \frac{\tilde{r} }{2 \sqrt{a}\,\sinh(\pi \tilde{r}) }  \frac{\sqrt{q_L'(-3a)}}{\sqrt{\kappa-\kappa_{\rm min}}}\, .
\end{equation}
Using \eref{eq:hqi} in \eqref{eq:bks1}, as $\kappa \to \kappa_{\rm min}$, we get, 
\begin{equation}
\mspace{-20mu}
\label{eq:Hkmin}
    H(\kappa) \simeq 
     \frac{A_1}{\sqrt{\kappa-\kappa_{\rm min}}}\,; 
\, \,
   A_1= \frac{\tilde{r} }{2 \sqrt{a}\,\sinh(\pi \tilde{r}) } \, \frac{   (\sigma\sqrt{\pi})^{1/2} }{\sqrt{\exp\left({-\frac{(L-3a)^2}{\sigma ^2}}\right)- \exp\left({-\frac{(L+3a)^2}{\sigma ^2}}\right)}}\,.
\end{equation}

On the other hand, as $\kappa \to \kappa^*$ from below, $H(\kappa)$ approaches a constant value $H(\kappa^*)$. Using  \eref{eq:bks1} and \eref{eq:hu3} we get
\begin{equation}
\label{eq:bks1kappas}
H(\kappa^*) =  \frac{   \sigma\sqrt{\pi} }{\exp\left({-\frac{(L-a)^2}{\sigma ^2}}\right)- \exp\left({-\frac{(L+a)^2}{\sigma ^2}}\right)}\,\frac{\tilde{r}}{4 a\,\sinh(\pi \tilde{r}/2) } \equiv A_2 \, .
\end{equation}

\subsubsection{Regime II: $\kappa^* \leq \kappa \leq \kappa_{\rm max}$: }
\label{subsubk2}

In this regime, where $u\in [-a,a]$,  there are two roots of $q_L(u) = \kappa$. Adding the contributions from both roots, we get 
\begin{align}
\label{eq:bkb}
H(\kappa)& =   \int_{-a}^{a} du\, h(u)\, 
  \delta[\kappa - q_L(u)] \nonumber \\
  &= \int_{-a}^{0} du\, h(u) \frac{\delta(u-q_L^{-1}(k))}{|q_L'(u)|}  +  \int_{0}^{a} du\, h(u) \frac{\delta(u-q_L^{-1}(k))}{|q_L'(u)|} \, ,
\end{align}
where $q_L^{-1}(\kappa)$ is the inverse of the function given in \eqref{eq:qL} in the respective ranges $[-a,0]$ and $[0,a]$. Since $q_L(u)$ is symmetric about $u=0$ [see \fref{fig:qlu} and equation~\eref{eq:qL}], if $u  = q_L^{-1}(\kappa)$ denotes the inverse function in $[0,a]$, then the the inverse function in the range $[-a,0]$ is simply $-u$. Therefore, we get 
\begin{equation}
\label{eq:bks2}
H(\kappa)=  \frac{   \sigma\sqrt{\pi}\, \left[h\bigl(q_L^{-1}(k)\bigr) + h\bigl(-q_L^{-1}(k)\bigr) \right] }{\exp\left({-\frac{\bigl(L-q_L^{-1}(k)\bigr)^2}{\sigma ^2}}\right)- \exp\left({-\frac{\bigl(L+q_L^{-1}(k)\bigr)^2}{\sigma ^2}}\right)}\, \quad \text{where}~~\kappa^*\leq \kappa \leq \kappa_{\rm max},
\end{equation}
with $q_L^{-1}(k) \in [0,a]$ being the inverse of the function $q_L(u)$ given in \eref{eq:qL} in the range $u\in [0,a]$. 

We now analyze $H(\kappa)$ given in \eref{eq:bks2} near the two edges $\kappa^{*}$ (from above) and $\kappa_{\rm max}$.  At $\kappa =\kappa^*$, the two roots of the inverse function $q_L^{-1}(\kappa^*)$ are $-a$ and $a$. We recall the limiting behavior of the function $h(u)$ near $u=\pm a$  in \eref{eq:hlim}, i.e., $h(u)$ diverges as $1/\sqrt{a-u}$ as $u\to a$, whereas $h(-a)$ is a constant. Expanding $q_L(u)$ in Taylor series near $u=a$, we get 
\begin{equation}
\label{eq:amu}
    a-u = \frac{\kappa-\kappa^*}{|q_L'(a)|}\quad\text{where}~~ q_L(u)=
    \kappa~~\text{and}~~ q_L(a) = \kappa^*.
\end{equation}
Therefore using \eref{eq:amu} in \eref{eq:hlim}, near $q_L^{-1}(k)= a$, we get 
\begin{equation}
\label{eq:hqii}
    h\bigl(q_L^{-1}(k)\bigr) \simeq  \frac{\tilde{r} }{2 \sqrt{a}\,\sinh(\pi \tilde{r}) }  \frac{\sqrt{|q_L'(a)|}}{\sqrt{\kappa-\kappa^*}}\, .
\end{equation}
Using \eref{eq:hqii} in \eqref{eq:bks2}, and $q_L'(a) = -q_L'(-a)$, as $\kappa \to \kappa^*$ from above, we get, 
\begin{equation}
\label{eq:Hkabove}
    H(\kappa) \simeq 
     \frac{A_3}{\sqrt{\kappa-\kappa^*}}\, ; \, \, A_3=
      \frac{\tilde{r} }{2 \sqrt{a}\,\sinh(\pi \tilde{r}) } \, \frac{   (\sigma\sqrt{\pi})^{1/2} }{\sqrt{\exp\left({-\frac{(L-a)^2}{\sigma ^2}}\right)- \exp\left({-\frac{(L+a)^2}{\sigma ^2}}\right)}}.
\end{equation}
Therefore, the scaling function $H(\kappa)$ has a discontinuity at $\kappa=\kappa^*$, namely, $H(\kappa)$ approaches a constant given by \eref{eq:bks1kappas} as $\kappa\to \kappa^*$ from below whereas it has a square-root divergence $H(\kappa) \sim 1/\sqrt{\kappa-\kappa^*}$ as $\kappa\to \kappa^*$ from above as given in \eref{eq:Hkabove}. 

Finally, to analyze $H(\kappa)$ given in \eref{eq:bks2} near $\kappa_{\rm max}$, we note that near $u=q_L^{-1}(\kappa_{\rm max}) =0$, the numerator of \eref{eq:bks2} approaches a constant whereas the denominator diverges linearly in $u = q_L^{-1}(\kappa)$.
Expanding $q_L(u)$ in Taylor series in \eref{eq:qL} around $u=0$, we have $q_L(u)=q_L(0)-(u^2/2) |q_L''(0)| +\dotsb$. Therefore, as $\kappa\to \kappa_{\rm max}$ from above, we have 
\begin{equation}
\label{eq:unear0}
    u \equiv  q_L^{-1}(\kappa) \simeq \frac{\sqrt{2 (\kappa_{\rm max} -\kappa)}}{\sqrt{|q_L''(0)|}}\quad\text{where}~~ \kappa_{\rm max} = q_L(0)~~\text{and}~~ q_L(u)=\kappa, 
\end{equation}
and from \eref{eq:qL},  $|q_L''(0)|=4 L e^{-L^2/\sigma ^2}/(\sqrt{\pi } \sigma ^3)$.
Therefore, expanding the denominator of \eref{eq:bks2} in Taylor series, using \eref{eq:unear0}, and $h(0)$ from \eref{eq:hu3}, we get
\begin{equation}
\label{eq:Hkmax}
    H(\kappa) \simeq \frac{A_4}{\sqrt{\kappa_{\rm max}-\kappa}}; \, \, A_4=\frac{\sigma^{3/2} \pi^{1/4}  \,e^{L^2/(2\sigma^2)} }{\sqrt{2L} } \,  \frac{\tilde{r}\, \cosh(2\pi \tilde{r}/3)}{\sqrt{3}\, a\, \sinh(\pi \tilde{r})}.
\end{equation}

\begin{figure}
    \centering
    \includegraphics[width=0.5\linewidth]{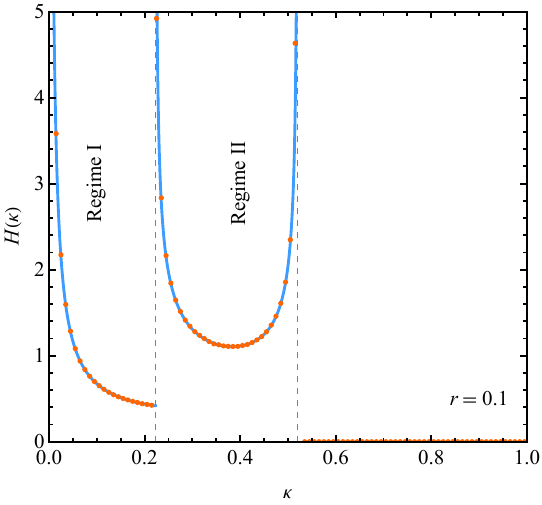}~~\includegraphics[width=0.5\linewidth]{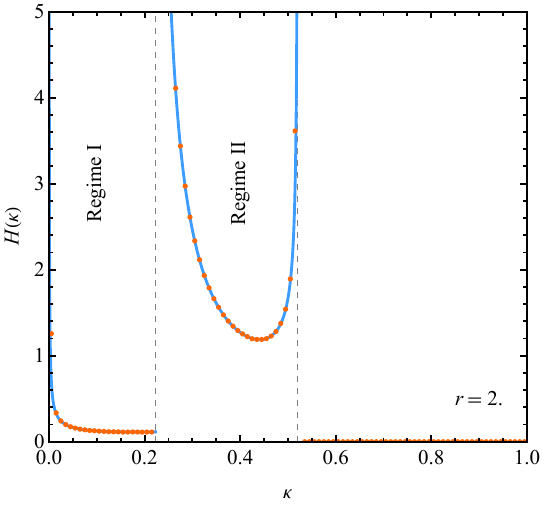}
    \caption{The scaling function $H(\kappa)$ describing the FCS in \eref{FCS-scaling} is plotted against $\kappa$ for $r=0.1$ (left) and $r=2.0$ (right) with $L=0.5$. The solid lines plot the analytical solutions given in \eref{eq:bks1} and \eref{eq:bks2} in the two regimes I and II, respectively.  The points are obtained from numerical simulations with $35\times 10^6$ realizations and $N=10^6$. We set $\omega=1$, $\sigma=1$ and $a=1$. The two dashed vertical lines indicate the positions of $\kappa^*=0.22\dots$ and $\kappa_{\rm max}=0.52\dots$ respectively. The nonzero position of $\kappa_{\rm min}= 0.0002\dots$ is not visible in the figures. One observes the divergence of $H(\kappa)$ as $\kappa\to \kappa_{\rm min}$,  $\kappa\to \kappa^*$ (from the right) and $\kappa\to \kappa_{\rm max}$ and the fact that $H(\kappa)\to \text{const.}$  as $\kappa\to \kappa^*$ from the left.}
    \label{fig:fcs}
\end{figure}

\noindent 
We now summarize the FCS discussed in the two regimes (I and II) studied in  \sref{subsubk1} and \sref{subsubk2}. Equation \eref{eq:bks1}  and \eref{eq:bks2} together give us the FCS. In \fref{fig:fcs}, we plot the FCS given in \eref{eq:bks1} and \eref{eq:bks2} and also demonstrate excellent agreement with numerics. 
The scaling function $H(\kappa)$ exhibits a square-root divergence at the two supports $\kappa_{\rm min}$ and $\kappa_{\rm max}$ as given in \eref{eq:Hkmin} and \eref{eq:Hkmax} respectively. Moreover, it has a discontinuity at an intermediate point $\kappa_{\rm min} < k^* < \kappa_{\rm max}$. As $\kappa\to \kappa^*$ from below, the scaling function $H(\kappa)$ approaches a constant given in \eref{eq:bks1kappas}. On the other hand, $H(\kappa)$ diverges as $1/\sqrt{\kappa-\kappa^*}$ as $\kappa\to \kappa^*$ from above, as given in \eref{eq:Hkabove}. Thus, summarizing,  
\begin{equation}
    H(\kappa) \simeq 
    \begin{cases}
    \displaystyle
        \frac{A_1}{\sqrt{\kappa-\kappa_{\rm min}}} &\quad\text{as}~~ \kappa\to \kappa_{\rm min}\\[5mm]
        A_2 &\quad\text{as}~~ \kappa\to \kappa^*~\text{from below/left}\\[5mm]
        \displaystyle
        \frac{A_3}{\sqrt{\kappa-\kappa^*}} &\quad\text{as}~~ \kappa\to \kappa^*~\text{from above/right}\\[5mm]
        \displaystyle
        \frac{A_4}{\sqrt{\kappa_{\rm max}- \kappa}} &\quad\text{as}~~ \kappa\to \kappa_{\rm max}
    \end{cases}
\end{equation}
 where the constants $A_1$, $A_2$, $A_3$, and $A_4$ are given in \eref{eq:Hkmin}, \eref{eq:bks1kappas}, \eref{eq:Hkabove} and \eref{eq:Hkmax} resepctively. 

\subsubsection{Generalizing FCS to arbitrary domain $[L_1,L_2]$:}
\label{subsubfcs}

\begin{figure}
    \centering
\includegraphics[width=0.8\linewidth]{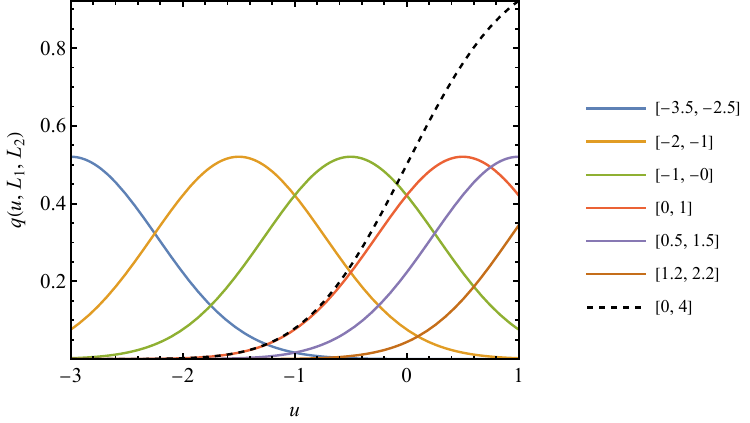}
    \caption{Plots of $\kappa = q(u,L_1,L2)$ given in \eref{eq:qL1L2} as a function of $u$, for various domains $[L_1,L_2]$ mentioned in the legend. The black dashed line is the special case close to the ``index problem", i.e., $L_1=0$ and $L_2\to \infty$.} 
    \label{fig:qlu2}
\end{figure}

Now, we generalize the FCS to arbitrary domain $[L_1,L_2]$. It is worth noting that the special case $L_1=0,L_2 \to \infty$ is referred to as the ``index problem".  For the arbitrary domain case, as in the FCS in \eref{FCS-scaling}, the probability of finding  $N_{L_1,L_2}$ number of particles in a domain $[L_1,L_2]$ is given by
\begin{equation}
  P(N_{L_1,L_2},N) \simeq \frac{1}{N} H\left(\frac{N_{L_1,L_2}}{N}\right)\quad\text{with}~~  H(\kappa)=   \int_{-3a}^{a} du\, h(u)\, 
  \delta[\kappa - q(u,L_1,L_2)],
  \label{FCS-scaling2}
\end{equation}
where $q(u,L_1,L2)$ is given by [generalization of the special case in \eref{eq:qL}] 
\begin{equation}
\label{eq:qL1L2}
q(u,L_1,L_2)=\int_{L_1}^{L_2} \frac{1}{\sqrt{\pi \sigma^2}}\,e^{-\frac{1}{\sigma^2}(x_j-u)^2}\, dx = 
\frac{1}{2} \left[\text{erf}\left(\frac{u-L_1}{\sigma }\right)+\text{erf}\left(\frac{L_2-u}{\sigma }\right)\right]\, .
\end{equation}
As before, in order to carry out the integral in \eref{FCS-scaling2}, we need to invert the function $q(u,L_1,L_2)$ in \eref{eq:qL1L2}. It is useful to elucidate $q(u,L_1,L_2)$ as a function of $u$ for different domains of $[L_1,L_2]$. As evident from \fref{fig:qlu2}, depending on the domain, the function $q(u,L_1,L_2)$  is either single-valued or multi-valued in the range $u \in [-3a, a]$. Subsequently, the function can be inverted as described above for the symmetric domain $[-L,L]$. 

Finally, we note that for in the ``index problem", setting $L_1=0$ and $L_2 \to \infty$ in \eqref{eq:qL1L2}, we get $q(u,0,\infty) =  \bigl[1+\text{erf}(u/\sigma)\bigr]\big/2$ which turns out to be a single valued function. Therefore, from \eref{FCS-scaling2}, the scaling function for the distribution of the index fraction $N_{0,\infty}/N$ is given by,
\begin{equation}
\label{eq:index}
    H(\kappa) = \sqrt{\pi}\,\sigma \,e^{u(\kappa)^2/\sigma^2}\, h\bigl(u(\kappa)\bigr), \quad\text{where}~~ u(\kappa) = \sigma \,\mathrm{erf}^{-1}(2\kappa-1),
\end{equation}
with $\mathrm{erf}^{-1}$ being the inverse error function, i.e., $\mathrm{erf}[\mathrm{erf}^{-1}(z)] = \mathrm{erf}^{-1}[\mathrm{erf} (z)] =z $ and $\mathrm{erf}^{-1}(z)= - \mathrm{erf}^{-1}(-z)$. The scaling function $H(\kappa)$ in \eref{eq:index} for the index distribution in supported in the finite range $\kappa\in [(1-\mathrm{erf}(3a/\sigma))/2, (1+\mathrm{erf}(a/\sigma))/2]$.

\section{Numerical simulation procedure}
\label{sec:simulation}

In our numerical simulations, we draw the random variables directly from the stationary JPDF~\eref{eq:ren_NESS1U}. To generate the set of correlated random variables with a common random mean $u$, we follow the steps below for each realization of the simulation.
\begin{enumerate}
    \item We first draw a random time $\tau$ from the exponential distribution $r\,e^{-r\tau}$.

    \item We then construct another random variable $u$ from $\tau$ using \eref{eq:umuR}.

    \item Next, we draw $N$ Gaussian random variables $\{x_1, x_2, \dots, x_N\}$ independently with a common mean $u$ and variance $\sigma^2$.
\end{enumerate}
We repeat steps (i)--(iii) to generate an ensemble of random variables according to the JPDF given in~\eref{eq:ren_NESS1U}. We use this ensemble of random variables to obtain the statistics of the observables of interest discussed in sections \ref{subsec:dp}---\ref{sub:fcs}.  \Fref{fig:hu}  verifies that the PDF of the random mean $u$ generated by using a large number of realizations of steps (i) and (ii) indeed agrees with  $h(u)$ given in~\eref{eq:hu3}.

\section{Summary and Outlook}
\label{sec:conc}

To summarize, we investigated a system of $N$ noninteracting bosons in one dimension that are simultaneously subjected to resetting with a rate $r$. The initial state is the ground state of a harmonic oscillator centered around a position ($x = a$) and the time evolution is with a different Hamiltonian, more precisely, an oscillator centered around another location ($x=-a$). 
The subsequent unitary dynamics is interrupted by simultaneously resetting all the bosons back
to the initial state, which results in emergent strong attractive correlations. 
We showed that the system reaches a nonequilibrium stationary state with 
a joint distribution that is non-factorizable. We demonstrated that the stationary joint distribution has a conditionally independent
and identically distributed (CIID) structure, given in \eref{eq:Psit2_gauss},  which we further exploited to compute several observables analytically. 
The density profile in \fref{fig:dens} shows an interesting bimodal to unimodel transition. In \fref{fig:time} we show the quantum dynamics of the density profile and its eventual approach to steady state in the presence of resetting. 
To unravel the effects of these correlations, we computed the two-point correlation functions [equation \eref{Cij}], order statistics [\fref{fig:os}], and the full counting statistics [\fref{fig:fcs}]. The results of these quantities are very distinct from that of uncorrelated or weakly correlated bosons indicating the dynamical emergence of a strongly correlated Bose gas. In particular, the FCS exhibits rather surprising features. First, in the thermodynamic limit, the lower support $\kappa_{\rm min}$ of the scaling function describing the FCS is strictly greater than zero, and the upper support $\kappa_{\rm max}$ is strictly less than unity. This indicates that a given region $[-L, L]$ can neither be completely empty nor completely full. The second surprising feature is that there is a discontinuity of the scaling function at an intermediate fraction $\kappa_{\rm min} < \kappa^* < \kappa_{\rm max}$. The scaling function diverges as $\kappa\to\kappa^*$ from above, whereas it approaches a constant when $\kappa$ approaches $\kappa^*$ from below.  

The focus of this work was on noninteracting bosons in a harmonic trap subjected to quantum resetting. There are many new interesting future directions one can pursue. 
For example, it will be interesting to extend our study to trapped noninteracting fermions subjected to a similar quantum resetting protocol where the Pauli exclusion principle will play an important role.  It will also be interesting to consider a protocol where the center of the harmonic trap is repeatedly switched back and forth~\cite{SM24} instead of the strict instantaneous reset protocol considered in this work.  Realizing our results in 
experiments will be interesting.
In this work, we restricted ourselves to noninteracting bosons. It would be further interesting to explore quantum resetting in interacting systems where the correlations between two particles in the nonequilibrium stationary state has two origins: (i) due to the inherent interactions between the particles and (ii) generated dynamically by the simultaneous quantum resetting. It will be interesting to study the combined effects of (i) and (ii).

\section{Acknowledgements}
We thank G. Schehr for useful discussions. M. K. acknowledges the support of the Department of Atomic Energy, Government of India, under project no. RTI4001. M. K. thanks the VAJRA faculty scheme (No. VJR/2019/000079) from the Science and Engineering Research Board (SERB), Department of Science and Technology, Government of India. S. N. M. and S. S. acknowledges the support from the Science and Engineering Research Board (SERB, Government of India), under the VAJRA faculty scheme (No. VJR/2017/000110). M. K. and S. S. thank the hospitality of Laboratoire de Physique Théorique et Modèles Statistiques (LPTMS), University Paris-Saclay where a major part of the work took place. We thank the support from the International Research Project (IRP) titled `Classical and quantum dynamics in out of equilibrium systems' by CNRS, France. M. K. thanks the hospitality of Raman Research Institute, Bangalore.
M. K. and S. S. thank the hospitality of Laboratoire de Physique Théorique et Hautes Energies (LPTHE), Sorbonne Université, Paris, France.

\end{document}